# Very low effective Schottky barrier height for ErSi$_{2-x}$ contacts on *n*-Si through arsenic segregation

Nicolas Reckinger,[1,a)] Claude Poleunis,[2] Emmanuel Dubois,[3] Constantin Augustin Duțu,[1] Xiaohui Tang (唐晓慧),[1] Arnaud Delcorte,[2] and Jean-Pierre Raskin[1]

[1]*ICTEAM institute, Université catholique de Louvain, Place du Levant 3, 1348 Louvain-la-Neuve, Belgium*

[2]*IMCN institute, Surface characterisation platform (ToF-SIMS), Université catholique de Louvain, Croix du Sud 1, 1348 Louvain-la-Neuve, Belgium*

[3]*IEMN/ISEN, UMR CNRS 8520, Avenue Poincaré, Cité Scientifique, 59652 Villeneuve d'Ascq Cedex, France*

The segregation of As$^+$ ions implanted into thin Er films deposited on *n*-Si substrates is studied after ErSi$_{2-x}$ formation. The same lowering of the effective Schottky barrier height (SBH) below 0.12 eV is obtained at moderate annealing temperatures, regardless of the redistribution of As dopants at the ErSi$_{2-x}$/Si interface. On the other hand, if the implanted dose is slightly enhanced, the annealing temperature required to reach sub-0.12-eV effective SBH can be further reduced. This process enables the formation of very low effective SBH ErSi$_{2-x}$/*n*-Si contacts with a low thermal budget.

---

[a)] Electronic mail : nicolas.reckinger@uclouvain.be.

In the relentless race to miniaturization of Si-based transistors, the Schottky barrier metal-oxide-semiconductor field-effect transistor (SBMOSFET) appears as a promising candidate.[1] The key parameter governing the current injection between the source/drain silicide contacts and the low doping channel is the Schottky barrier height (SBH). To facilitate the carrier transport through such devices, the SBH must preferably be as low as possible. Simulations showed indeed that SBMOSFETs can compete with traditional MOSFETs if the SBH is less than 0.1 eV.[2] Most of silicides present a high SBH $\Phi_B$ for both electrons and holes. However, only a few of them feature band-edge properties meaning that the Fermi level is strongly pinned close to either the valence or the conduction band. Low SBHs to holes can be obtained with PtSi ($\Phi_{Bp} \approx 0.15\,\text{eV}$)[3] or IrSi ($\Phi_{Bp} \approx 0.11\,\text{eV}$),[4] and low SBHs to electrons can be obtained from silicides based on rare-earth (RE) elements, like Er or Yb ($\Phi_{Bn} \approx 0.28\,\text{eV}$).[5] To reach the sub-0.1-eV SBH target, the segregation technique, proposed three decades ago by Thornton,[6] was recently revived by Kinoshita *et al.*.[7] Impurity segregation consists in piling up various kinds of impurities (dopants,[8,9,10] valence mending adsorbates,[11,12,13] or others[14,15]) at the silicide/Si interface. In this way, the SB width is shrunk, thereby promoting carrier injection by tunneling and resulting in a lower effective SBH ($\Phi_B^{\text{eff}}$). Segregation was extensively investigated for midgap silicides like NiSi[8,10,11,12,14,15] and valence-edge silicides like PtSi[8,9,12,13] with various types of impurities. One group studied dopant segregation (DS) implemented to RE silicides.[16] In the present article, we explore As segregation associated to a conduction-band-edge silicide, namely, Er disilicide (ErSi$_{2-x}$). We demonstrate a very efficient $\Phi_B^{\text{eff}}$ reduction correlated to As piling-up at the ErSi$_{2-x}$/Si interface.

The starting substrates are *n*-type lowly doped bulk Si(100) wafers (phosphorus-doped with a resistivity of 5-10 Ω cm). First, a thick layer of thermal SiO$_2$ is grown. Next, photolithography and buffered hydrofluoric (HF) acid etch are used to pattern face-to-face Schottky diodes for $\Phi_{Bn}^{\text{eff}}$ extraction. The device consists of two 1×1 mm$^2$ ErSi$_{2-x}$ contacts separated by a 100 μm wide Si gap. After photoresist removal and just before the insertion of the wafer into the evaporator, HF dip is carried out to remove native Si oxide. Then, 25 nm of Er capped by 10 nm of Ti are deposited by e-beam under high vacuum conditions on three wafers at the same time. The wafers are then implanted at 25 keV with As$^+$ ions, each with a different dose (5×10$^{14}$, 10$^{15}$, and 5×10$^{15}$ cm$^{-2}$). The implantation energy is determined by simulation so as to maximize and confine the dopants into the Er film and the Si layer consumed during the ErSi$_{2-x}$ growth. Afterwards, the ErSi$_{2-x}$ contacts are grown by rapid thermal annealing from 400 to 700 °C with a step of 100 °C. Finally, the capping layer and unreacted Er are both stripped.

First, it is necessary to verify if the implantation affects the formation of ErSi$_{2-x}$ in some way. To that purpose, θ-2θ x-ray diffraction (XRD) analyses are performed on the ErSi$_{2-x}$ films annealed at 400 and 600 °C for the three doses. The corresponding spectra are exposed in Fig. 1. A peak at 2θ = 27.2°, related to ErSi$_{2-x}$, is recorded for all the samples. Still, the peak intensity is much higher for the samples annealed at 600 °C compared to 400 °C. The same analysis conducted on unimplanted samples evidences a minor difference only between 400 and 600 °C (not shown). In addition, it plainly appears that the ErSi$_{2-x}$ peak for the dose of 5×10$^{15}$ cm$^{-2}$ is much weaker, more strikingly at 600 °C. The previous results thus indicate that (1) implantation at very high dose considerably alters the crystallinity of ErSi$_{2-x}$ (that is plausibly amorphized) and (2) implantation retards the formation of ErSi$_{2-x}$.

Figure 2(a) shows the experimental Arrhenius plots (AP) corresponding to the dose of 5×10$^{14}$ cm$^{-2}$ for all the annealing temperatures and for a reference sample without As$^+$ implantation. The APs are obtained from temperature-dependent current-voltage (*I-V-T*) measurements performed on face-to-face Schottky diodes between 90 and 290 K. Only the curves corresponding to $V$ = 1 V are displayed for the sake of readability. Compared to the reference sample ($\Phi_{Bn}$ = 0.3 eV), it can be noted that the electrical characteristics are all affected, even at 400 °C. For a given SBH, the two-contact system is ohmic at high temperature (positive slope in the AP) while it becomes rectifying at low temperature (negative slope in the AP). A useful empirical parameter to obtain a first qualitative estimate of the efficiency of DS to lower $\Phi_{Bn}$ is the transition temperature ($T_{tr}$) between both transport regimes: the lower $T_{tr}$, the lower $\Phi_B^{eff}$. As a result, the efficiency of DS can be roughly determined by the temperature corresponding to the inflexion point of the AP for a fixed $V$. In the case of Fig. 2(a), $T_{tr}$ is observed to progressively drop between 400 and 500 °C and is no more visible for 600 and 700 °C (at least above 90 K), reflecting a concomitant lowering of $\Phi_{Bn}^{eff}$. To be more quantitative, $\Phi_{Bn}^{eff}$ is extracted as previously described.[3,5] It is determined to amount to ~0.25 and ~0.18 eV for 400 and 500 °C, respectively. Since $T_{tr}$ is less than 90 K for 600 and 700 °C, it is only possible to provide an upper bound for $\Phi_{Bn}^{eff}$ which is evaluated to ~0.12 eV. Nevertheless, the analysis can be further refined by comparing the *I-V* characteristics at 90 K, as depicted in Fig. 2(b). For identical Si series resistances, the current at $V$ = 1 V is higher for the 700 °C sample, indicative of enhanced electron injection by tunneling and, as a result, of a lower $\Phi_{Bn}^{eff}$. In addition, it is worth noting that the *I-V* characteristics are both ohmic. Remarkably, for the doses of $10^{15}$ and $5\times10^{15}$ cm$^{-2}$, the *I-V* characteristics are ohmic for all the annealing temperatures down to 90 K (not shown) and the corresponding $\Phi_{Bn}^{eff}$ is in consequence also found to amount to ~0.12 eV at most.

Figure 2(c) features the normalized APs for the same samples as in Fig. 1(b). Because these samples possess different silicon series resistances, contrary to the samples of Fig. 2(a), the curves are normalized by $I/T^2|_{T=290K}$ for a fair mutual comparison. It can be discerned that, for a fixed *T*, *I* grows with a higher dose, more especially at low *T*. The ratio rises by 26% upon doubling the implanted dose (between $5\times10^{14}$ and $10^{15}$ cm$^{-2}$), while the discrepancy between the greatest doses is weak, with a 9.5% increment only, although the dose is multiplied by 5. It appears that increasing the dose beyond $10^{15}$ cm$^{-2}$ does not bring any significant improvement in tunneling current injection. Moreover, the degradation of the crystalline quality of ErSi$_{2-x}$ inferred from the XRD study does not seem to impinge upon $\Phi_{Bn}^{eff}$, contrary to previous findings for ErSi$_{2-x}$ without DS. The beneficial effect of DS appears thus predominant over the deterioration of the silicide crystallinity.

It is worth briefly discussing here the pertinence of the concept of effective SBH and its ensuing extraction procedure. The effective SBH reflects the dependence of the SB profile on the electric field that results in an apparent lowering of the SBH. It is usually written as[17]

$$\Phi_{Bn}^{eff} = \Phi_{Bn} - \sqrt{\frac{qE_S}{4\pi\varepsilon}} - \alpha E_S,$$

with *q* the electric charge, $E_S$ the electric field at the silicide/Si interface, $\varepsilon$ the dielectric permittivity of Si, and $\alpha$ the tunneling distance. The $\sqrt{qE_S/4\pi\varepsilon}$ term accounts for the Schottky effect while $\alpha E_S$ represents the impact of tunneling injection. In the case of DS, this last term is by far the most contributing one. The sole consideration of a lower numerical value for $\Phi_{Bn}^{eff}$ in the thermionic-field emission model used to fit data is convenient but insufficient to properly predict the low *T* behavior of the electrical characteristics. In fact, a more complex mechanism, involving a SB profile thinning modulated by *V*, should be considered instead.[18] For that reason, the extracted $\Phi_{Bn}^{eff}$ should be regarded as a convenient empirical parameter to assess the efficiency of DS, rather than a physically meaningful one.

In order to examine the redistribution of As after silicidation and to correlate it to the electrical measurements, time-of-flight secondary ion mass spectrometry (ToF-SIMS) depth profiling is conducted on the samples presented in Fig. 1(b). The Cs$^+$ primary beam is operated at 1 keV with a target current of 130 nA and a sputtered area of 300×300 μm$^2$. The ejected material is analyzed with a 30 keV $Bi_3^+$ beam over a surface of 100×100 μm$^2$. The corresponding As and Si profiles are revealed in Fig. 3. The location of the ErSi$_{2-x}$/Si interface is marked by a dashed vertical line. That position differs slightly from one wafer (dose) to another due to small variations in the evaporated Ti and Er thicknesses. For each sample, we can discern the original peak of implanted As, near 100 s of sputter time, coinciding with the cap/ErSi$_{2-x}$ interface. Expectedly, the higher the implanted As$^+$ dose, the more the accumulated dopants at the ErSi$_{2-x}$/Si interface. A very distinct As pileup occurs for the dose of 5×10$^{15}$ cm$^{-2}$, apparently in the ErSi$_{2-x}$ film, in close vicinity of the interface. For the weakest doses, no clear As accumulation is observed. In spite of unequivocal detection of As piling-up, the corresponding two APs turn out to be significantly influenced by DS. Furthermore, although the ToF-SIMS profile and the interfacial concentration are very different for each dose, the impact on the electrical behavior appears to be similar (at least, as can be deduced in the considered range of measurement temperatures). DS is usually attributed to the interplay between, on the one side, the creation of a dipole at the interface with Si owing to activated impurities and, on the other side, the modification of the silicide work function at the same interface due the formation of new chemical bonds with the implanted species. Here, it appears that the total As concentration segregated at the ErSi$_{2-x}$/Si interface weakly influences $\Phi_{Bn}^{eff}$, sign that the physical mechanism at work is confined to the interface and is more likely the consequence of dipole formation. In that case, the magnitude of $\Phi_{Bn}^{eff}$ is rather related to the fraction of activated dopants in Si at the very close vicinity to the interface. Since it is observed that tunneling injection saturates for a dose superior to 10$^{15}$ cm$^{-2}$, it is invoked that the threshold of maximum activated dopant concentration is nearly attained due to the limited As solid solubility in Si (~8×10$^{19}$ cm$^{-3}$ at 700 °C[19]).

In summary, the impact of As segregation on the effective SBH of ErSi$_{2-x}$/$n$-Si contacts grown between 400 and 700 °C is investigated. XRD highlights that implantation retards the growth of ErSi$_{2-x}$. For a moderate implantation dose of 5×10$^{14}$ cm$^{-2}$, the extracted $\Phi_{Bn}^{eff}$ turns out to drop from ~0.25 eV at 400 °C down to at most ~0.12 eV beyond 600 °C. If the implanted dose is higher (≥ 10$^{15}$ cm$^{-2}$), $\Phi_{Bn}^{eff}$ lies below ~0.12 eV, even at 400 °C. However, implantation at very high dose hardly improves the tunneling injection, probably due to saturating dopant activation and it is also detrimental to the crystallinity of ErSi$_{2-x}$. ToF-SIMS depth profiling evidences As pileup for the highest dose while it is barely perceptible for the lower doses ($5\times10^{14}$ and $10^{15}$ cm$^{-2}$), despite the conspicuous effect of DS on all the electrical characteristics.

This work is supported by the European Commission through the NANOSIL network of excellence (FP7-IST-NoE-216171). The authors acknowledge the financial support of National Foundation for Scientific Research (F.N.R.S. - Belgium) for ToF-SIMS acquisition.

[1] E. Dubois and G. Larrieu, Solid-State Electron. **46**, 997 (2002).

[2] D. Connelly, C. Faulkner, and D. E. Grupp, IEEE Electron Device Lett. **24**, 411 (2003).

[3]E. Dubois and G. Larrieu, J. Appl. Phys. **96**, 729 (2004).

[4]G. Larrieu, E. Dubois, X. Wallart, and J. Katcki, J. Appl. Phys. **102**, 094504 (2007).

[5]N. Reckinger, X. Tang, V. Bayot, D. A. Yarekha, E. Dubois, S. Godey, X. Wallart, G. Larrieu, A. Łaszcz, J. Ratajczak, P. J. Jacques, and J.-P. Raskin, Appl. Phys. Lett. **94**, 191913 (2009).

[6]R. L. Thornton, Electron. Lett. **17**, 485(1981).

[7]A. Kinoshita, Y. Tsuchiya, A. Yagishita, K. Uchida, and J. Koga, Symposium on VLSI Technology **2004**, 168.

[8]Z. Qiu, Z. Zhang, M. Östling, and S.-L. Zhang, IEEE Trans. Electron Devices **55**, 396 (2008).

[9]G. Larrieu, E. Dubois, R. Valentin, N. Breil, F. Danneville, G. Dambrine, J.-P. Raskin, and J.-C. Pesant, Tech. Dig. - Int. Electron Devices Meet. **2007**, 147.

[10]S. F. Feste, J. Knoch, D. Buca, Q. T. Zhao, U. Breuer, and S. Mantl, J. Appl. Phys. **107**, 044510 (2010).

[11]Q. T. Zhao, U. Breuer, E. Rije, St. Lenk, and S. Mantl, Appl. Phys. Lett. **86**, 062108 (2005).

[12]H.-S. Wong, L. Chan, G. Samudra, and Y.-C. Yeo, Appl. Phys. Lett. **93**, 072103 (2008).

[13]E. Alptekin, M. C. Ozturk and V. Misra, IEEE Electron Device Lett. **30**, 331 (2009).

[14]M. Sinha, E. F. Chor and Y.-C. Yeo, Appl. Phys. Lett. **92**, 222114 (2008).

[15]W.-Y. Loh, H. Etienne, B. Coss, I. Ok, D. Turnbaugh, Y. Spiegel, F. Torregrosa, J. Banti, L. Roux, P.-Y. Hung, J. Oh, B. Sassman, K. Radar, P. Majhi, H.-H. Tseng, and R. Jammy, IEEE Electron Device Lett. **30**, 1140 (2009).

[16]G. Larrieu, D. A. Yarekha, E. Dubois, N. Breil, and O. Faynot, IEEE Electron Device Lett. **30**, 1266 (2009).

[17]J. M. Shannon, Appl. Phys. Lett. **24**, 369 (1974).

[18]N. Reckinger, X. Tang, E. Dubois, G. Larrieu, D. Flandre, J.-P. Raskin, and A. Afzalian, Appl. Phys. Lett. **98**, 112102 (2011).

[19]V. E. Borisenko and S. G. Yudin, Phys. Status Solidi A **101**, 123 (1987).

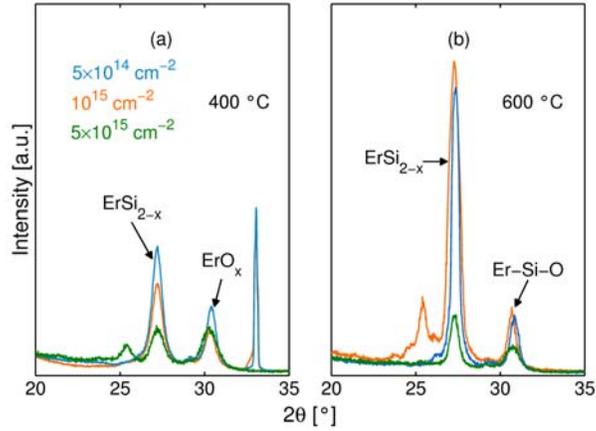

FIG.1. (Color online) XRD spectra for the samples annealed at 400 and 600 °C for each implanted dose. The same intensity scale is used in each graph, for direct comparison.

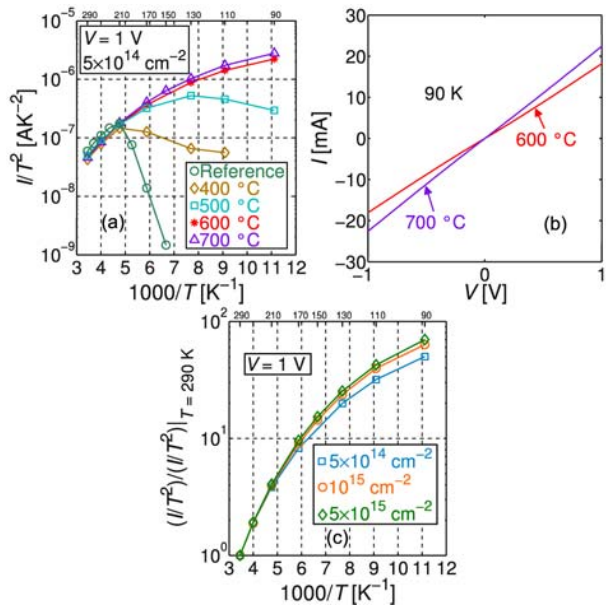

FIG.2. (Color online) (a) APs for the sample implanted at a dose of $5\times10^{14}$ cm$^{-2}$ for all the annealing temperatures, and for a reference sample without implantation. (b) *I-V* characteristics of the samples annealed at 600 and 700 °C at 90 K. (c) Normalized APs for the samples annealed at 600 °C for each implanted dose. The APs are displayed for $V = 1$ V.

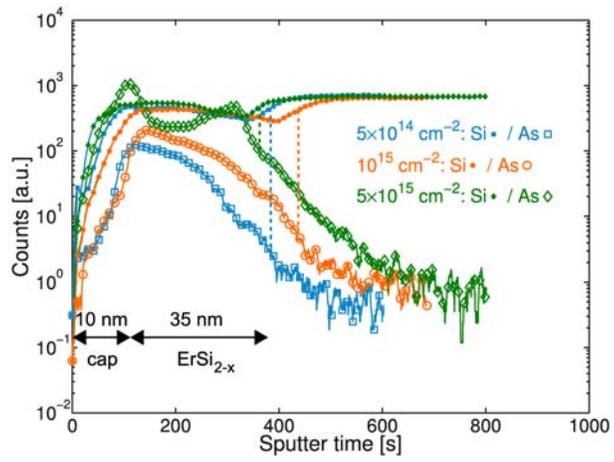

FIG.3. (Color online) ToF-SIMS Si and As depth profiles for the samples annealed at 600 °C for each implanted dose.